\documentclass[a4paper,twocolumn,showpacs,amsmath,amssymb,pre]{revtex4}
\usepackage{graphicx}
\usepackage{latexsym}

\begin{document}
 
\title{Fibonacci numbers in phyllotaxis : a simple model}

\author{Gilbert Zalczer}
\affiliation{Service de Physique de l'Etat Condens\'{e}, CEA Saclay, 91191 Gif-sur-Yvette cedex, France.}
\email{Gilbert.Zalczer@cea.fr}
 
\begin{abstract}{ A simple model is presented which explains the occurrence of high order Fibonacci number
parastichies in asteracae flowers by two distinct steps. First low order parastichies result from
the fact that a new floret, at its appearance is repelled by two former ones, then, in order to
accommodate for the increase of the radius, parastichies numbers have to evolve and can do
it only by applying the Fibonacci recurrence formula.
 } 
\end{abstract}
\maketitle 


The beautiful spirals observed on sunflowers or daisies have puzzled scientists since it was
realized that the numbers of these spirals belonged to the Fibonacci sequence. The basic
principle of phyllotaxis has been formulated as early as 1868 by Hofmeister \cite{Hofmeister1868} and states that
each new structure (leaf, floret, seed,etc.) appears "opposite" to the previous one - or ones.
For instance a new leaf often appears opposite to the previous one, which leads to an
alternate pattern. When however the new leaf feels the presence, with a smaller amplitude, of
the next previous one then the resulting pattern is starry as depicted in fig 1.
\begin{figure}[h]
\centering
\includegraphics 
{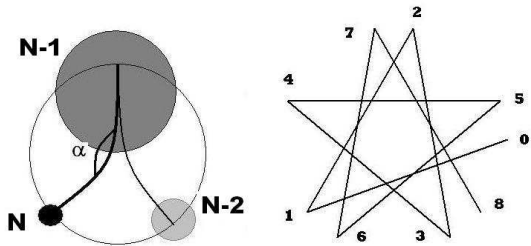}
\caption{ The new structure N is repelled by its predecessor N-1 and somewhat less by the former one
N-2. It appear at an angle $\alpha$ from its predecessor with $\alpha$ between $135  ^{\circ} $ and $144 ^{\circ} $. At steady
state, a starry pattern is formed where each structure is close to its 5th and 8th predecessors
and successors.
}
\end{figure}
 If leaves appear
in pairs, opposite to each other (decussate pattern), each pair is perpendicular to the previous
one. It is worth noting that alternate and spiral patterns can coexist in the same plant and that
a decussate pattern can turn into a spiral one on a stem: the pattern is sensitive to local
influences or noise.
In many observed cases and in simulations with a wide range of parameters the angle of the
starry pattern lies between $135 ^{\circ} =360  ^{\circ}\times 3/8 $ and $144 ^{\circ} =360  ^{\circ} \times 2/5 $  which means that the
"structure" of order n will be close to those of order $n \pm 8 $  and $n\pm 5$.If the growth is axial this
leads to the commonly observed pattern of crossing helices of order 5 and 8. A pineapple is a
typical example.
If the growth is radial, spirals (in place of helices) of higher order appear. These orders are not
random but belong to the Fibonacci sequence : this sequence is obtained by applying the
recurrence equation F(n+2)=F(n+1)+F(n) , starting with F(1)=F(2)=1. It starts as 1 1 2 3 5 8 13
21 34 55 ...
Several investigators \cite{Douady1996,Smith2006} have expanded the principle into very elaborate models,
considering that the appearance of a new leaf is conditioned by a repulsive potential created
by all the previous ones, writing down explicit forms for the model potential, the growth law
and the conditions of appearance and computing or simulating the solutions. Subtle changes
in any parameter lead to subtle changes in the star angle which would lead to different
parastichies. An implicit hypothesis of such computations is that the potential propagates in a
perfectly homogeneous medium. A growing plant does not fulfil this condition and the noise
due to its heterogeneity would blur such small effects. A "sight range" of 2 is enough anyway.
More involved mechanisms have also been proposed \cite{Klar2002}.

In the case of radial growth, however a fact which has been apparently overlooked until now is that the parastichies order cannot
keep constant along the radius of the flower. The example of a sunflower is shown in fig 2.
\begin{figure}[h]
\centering
\includegraphics {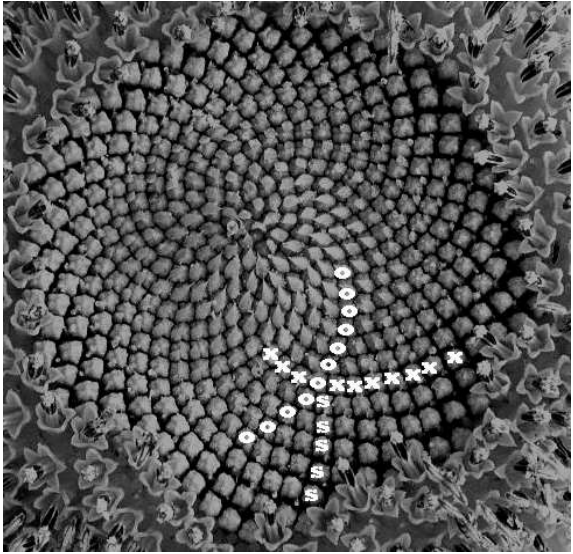}
\caption{The spiral of order 21('o') of this sunflower is stretched when the distance from the centre
increases and the clockwise next neighbour switches to the 's' spiral.}
\end{figure}
Let us consider the point at the intersection of
the 'o' spiral ( parastichy of order 21) and the 'x' spiral ( order 34). The next seeds along the 'o'
spiral are farther and farther away and no more in contact with each other. The obvious
clockwise spiral for the eye becomes the 's' one and interactions between neighbours, if any,
occur along it. Computing the new spiral order can be done by looking at fig 3.
\begin{figure}[h]
\centering
\includegraphics {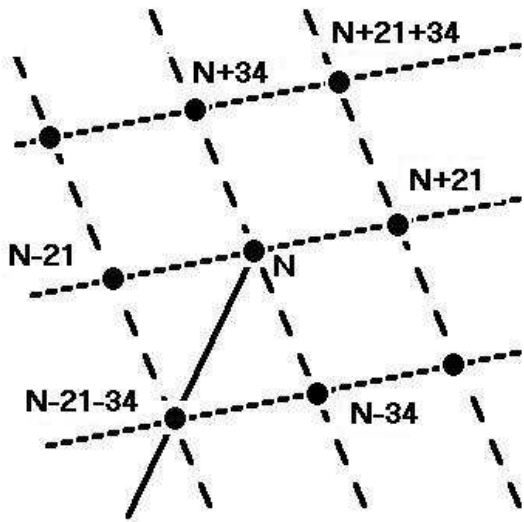}
\caption{Scheme of the crossing of spirals of order 21 and 34. When the N-21 point is pulled to the left,
the nearest neighbour of the N point in the lower left direction becomes N-21-34. The new
spiral order is obtained by following the Fibonacci recurrence formula. }
\end{figure}
The
appearance of successive seeds along a spiral of order n are separated by n-1 others in the
other spirals and their rank differ by n. Fig 3 outlines the spiral crossing at seed of rank N and
shows the rank of the neighbouring seeds. When the seed of rank N-21 is pulled far from the
seed N, the nearest neighbour of the latter becomes the predecessor of the former in its
order-34 spiral, which is precisely the application of the Fibonacci recurrence formula.
The occurrence of high order Fibonacci number parastichies can therefore be explained
simply by two distinct steps:
The place where each new floret appears is determined by a repulsion from the two previous
ones. This induces a starry distribution where a floret is very close to its 5th and 8th
predecessors and successors, leading to parastichies of order 5 and 8.
In case of radial growth, the spiral with lower order is stretched and becomes discontinuous
and is replaced by a new one whose order is the sum of the previous ones.
Parastichies obeying Lucas sequence have been observed occasionally. This sequence
obeys the same recurrence relation but a different couple of starting values.
The mechanism of the repulsion has still to be understood.It is believed to involve auxin
through complicated mechanisms \cite{Reinhardt2003,Prusinkiewicz2009}. It would be also interesting to investigate whether
some local interaction stabilises the spiral order, in contrast with some cactuses where a
vertical pattern is produced

\bibliography{fib}

\end{document}